# Unified Network Equation for Both Normal RLC Circuits and Josephson Junction Circuits

Yongliang Wang

*Abstract*—Josephson junction circuits, such as superconducting quantum interference devices (SQUIDs) and single-flux-quantum (SFQ) circuits, have been successfully applied in both analog and digital electronic domains. They are considered to be distinct from the normal resistor-inductor-capacitor (RLC) circuits, due to the superconducting physics. To bridge the gap between Josephson junction circuits and conventional RLC circuits, we introduce a magnetic-flux-generator (MFG) model to unify the definitions of Josephson junctions and normal noninductive components, and transform their circuits uniformly into MFG networks. A general network equation for both superconducting and normal MFG networks is derived. It shows that, Josephson junction circuits are the superconducting RLC networks modulated by Josephson currents; their analysis and simulation are unified with that of the normal RLC circuits by solving the general network equation.

## I. Introduction

JOSEPHSON junctions have created various integrated circuits used in both analog and digital domains [1]. Superconducting quantum interference devices (SQUIDs) are the typical Josephson junction circuits that have been widely used in magnetic field measurements [2], [3]; they are the flux-to-voltage converters with different structures [4], as shown in Fig. 1. Meanwhile, single-flux-quantum (SFQ) circuits [5] are the Josephson junction circuits used for digital computing systems. They are large-scale networks with Josephson junctions connected in parallel and serial; their typical circuit cell is similar to the SQUID circuit shown in Fig. 1(e). In practical applications, Josephson junction circuits are biased with external current sources and integrated with non-superconducting circuits, as shown in Fig. 2. They are actually the hybrid networks of Josephson junctions with normal resistor-inductor-capacitor (RLC) elements.

Different with the normal RLC circuits which are analyzed using currents and voltages as variables, the current-voltage relation of Josephson junctions and the flux-quantization law (FQL) for superconducting loops are both expressed with the macroscopic quantum phases of superconductors [6]. Since a nodal voltage $v_n$ is the derivative of its nodal phase $\varphi_n$ with respect to time, namely $d\varphi_n/dt = 2\pi v_n/\Phi_0$, according to the Josephson equations [7], the SPICE (simulation program with integrated circuit emphasis) tools for Josephson junction circuits, such as JSIM [8], JSPICE [9], PSCAN [10], or JoSIM [11], are developed by using the phase-based modified nodal analysis (MNA) method [12], in which the nodal voltages are replaced with the nodal phases [13].

However, two problems are confronted in the analysis of Josephson junction circuits using SPICE tools.

First, as the integration of nodal voltages with respect to time, nodal phases in the MNA method are the virtual variables rather than physical quantities for normal RLC elements and circuits. They are difficult to be understood by the electronic engineers who are trained with classic circuit theories. Meanwhile, the phase-based nodal analysis method cannot provide a general network equation for further dynamic studies, because the Josephson current is the sinusoidal function of the nodal phase difference between two terminals and cannot be expressed with the nodal admittance matrix. The MNA matrix used in SPICE tools is only the iteration formulas for calculating the numerical responses of Josephson junctions, and is not open to users.

Second, the circuit diagrams processed by SPICE tools are inconvenient to describe the phase variations inside branches. The vector potential **A** of the magnetic field $B$, $B = \nabla \times \mathbf{A}$, induces the gradient of the phase $\varphi(x,y,z,t)$ along the superconducting wires, namely $\nabla \varphi \approx \mathbf{A}$, and creates a phase difference $\Delta \varphi$ in branches with a certain length. The vector potential **A** is imposed by the external flux $\Phi_e$ and the branch currents through self or mutual inductances. Therefore, to describe the magnetic couplings between branches with conventional circuit diagrams, we should put phase sources in the branches to represent the effect of external flux $\Phi_e$, and add the current-controlled phase sources or multiport transformers to the branches to exhibit the mutual couplings between branches. For the Josephson junction circuits with a large number of branches, the circuit diagram may be too complicated to be analyzed, if we want to take all the magnetic couplings between branches into consideration.

To solve those problems, this article introduces a concept of magnetic-flux-generator (MFG) to unify the definitions of Josephson junctions and the normal noninductive components. Based on this concept, we can transform superconducting and non-superconducting circuits, as well as their hybrids, into MFG networks and derive a unified network equation to depict their dynamics. The general network equation of MFG networks bridge the gap between the Josephson junction circuits and normal RLC circuits. It shows that Josephson junction circuits are a kind of RLC circuits modulated by the

Yongliang Wang is with the State Key Laboratory of Functional Materials for Informatics, Shanghai Institute of Microsystem and Information Technology, Chinese Academy of Sciences (CAS), and the CAS Center for Excellence in Superconducting Electronics, Shanghai 200050, China (e-mail: wangyl@mail.sim.ac.cn).

Josephson currents inside Josephson junctions; their analyses have no differences with that of normal RLC circuits.

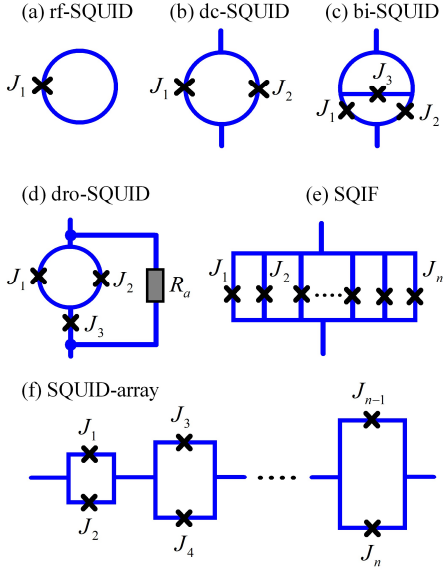

Fig. 1. Typical Josephson junction circuits, where a cross "×" denotes a Josephson junction: (a) radio-frequency (rf) SQUID; (b) direct-current (dc) SQUID; (c) bi-SQUID; (d) double-relaxation-oscillation (dro) SQUID, which is shunted with a normal resistor $R_a$; (e) superconducting quantum interference filter (SQIF); (f) SQUID array with dc-SQUIDs connected in serial.

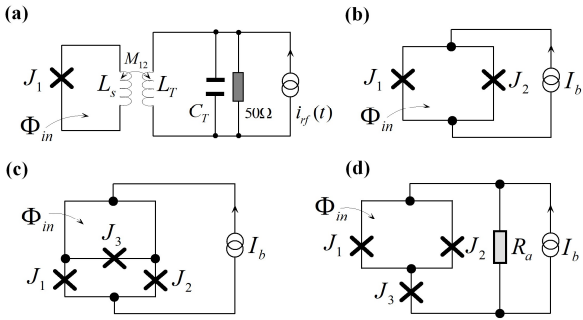

Fig. 2. SQUID circuits in practical operations: (a) Practical rf-SQUID circuit tuned by a rf current source $i_{rf}(t)$. (b) Current-biased dc-SQUID circuit. (c) Current-biased bi-SQUID circuit. (d) Current-biased dro-SQUID circuit.

## II. METHOD

### A. Magnetic-Flux-Generator Model

Josephson junction circuits and normal RLC circuits are a group of loops weaved by superconductor or perfect-conductor wires. Josephson junctions or normal RC components, such as a resistor, a capacitor, or their combination connected in parallel, are inserted between wires. They are the noninductive circuit components which dimensions are neglected.

A current-biased RC component inserted between two normal conductors implements the current-to-voltage conversion, when it is driven by a current source $i_b$ and the branch current $i_{EL}$, as shown in Fig. 3(a). Its equivalent circuit is illustrated in Fig. 3(b), where $u_{EL}$ is a potential difference between two terminals; $i_n$ is the current noise or current fluctuation, and $i_{RC}$ is the normal current flowing through the RC component under the voltage $u_{EL}$.

A current-biased RC component inserted between two superconductors implements the current-to-phase conversion, as shown in Fig. 3(c). It produces a quantum-phase difference $\theta_{EL}$ between two superconductive terminals. Its equivalent shown in Fig. 3(b) is same with the one of normal RC components.

The current-biased Josephson junction shown in Fig. 3(e) is actually a Josephson-current-shunted RC component inserted between two superconductors, according to the resistively-capacitively-shunted junction (RCSJ) model [14]. Its equivalent circuit shown in Fig. 3(f) has an additional Josephson current bypassing the normal RC component, compared to the equivalent circuit shown in Fig. 3(d).

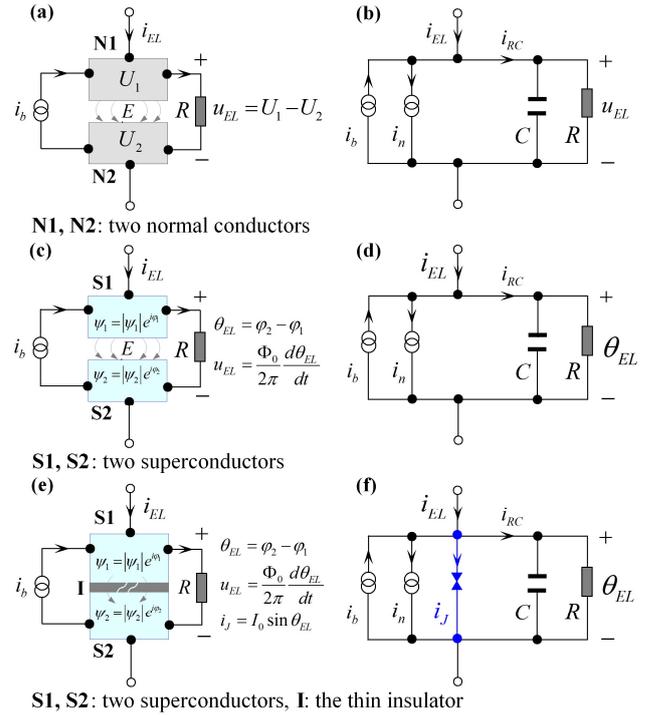

Fig. 3. Three typical types of current-biased RC components: (a) The current-biased RC component inserted between two normal conductor wires, where $U_1$ and $U_2$ are the electric potentials; (b) the equivalent circuit which is a current-biased network with a resistor and a capacitor connected in parallel. (c) The current-biased RC component inserted between two superconductor wires, where $\varphi_1$ and $\varphi_2$ are the macroscopic quantum phases; its (d) equivalent circuit. (e) The current-biased Josephson junction, and the (f) equivalent circuit which has an additional Josephson current $i_J$ tunneling through.

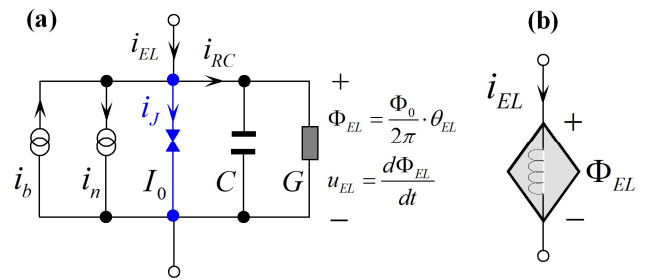

Fig. 4. (a) General equivalent circuit of MFGs, where $G = 1/R$ is the conductance of $R$. (b) Common circuit symbol of MFGs.

Three types of RC components are uniformly modeled as the magnetic-flux-generator (MFG) which equivalent circuit and electric symbol are shown in Fig. 4(a) and (b) respectively. Normal RC components are the MFG with a $I_0 = 0$, while Josephson junctions are the MFG with a $I_0 \neq 0$. As its name implies, the MFG is an active electric component that implements the current-to-flux conversion inside circuit loops. Therefore, we introduce a nominal flux [15] $\Phi_{EL}$ to redefine the voltage and phase differences at two terminals of MFGs,

$$u_{EL} = \frac{d\Phi_{EL}}{dt}; \theta_{EL} = 2\pi \frac{\Phi_{EL}}{\Phi_0} \quad (1)$$

where, $\Phi_0 = 2.07 \times 10^{-15}$ Wb.

With this MFG model, we can redraw all the RC components with the uniform electric symbol shown in Fig. 4(b), and transform Josephson junction circuits and normal RLC circuit, as well as their hybrids, into MFG networks. The method of building the MFG networks for Josephson junction circuits is illustrated in Appendix.

In the MFG network diagrams shown in Fig. 10-13, MFGs are interfering each other inside magnetically coupled loops, while loops are the entities that bind all the MFGs. Therefore, we treat the self-inductance and mutual-inductance as the internal properties of loops to avoid showing them in MFG network diagrams.

### B. MFG Network Analysis

Loops and MFGs are the two types of objects that constitute the MFG networks. For a given MFG network, we can define $P$ independent loops, as shown in Fig. 5(a); each loop, namely Loop-$i$ ($1 \leq i \leq P$), contains a loop-current $i_{mi}$ which is defined according to the mesh analysis method [16]; its external applied flux is $\Phi_{ei}$, and the total flux coupled in the loop is $\Phi_{mi}$. Meanwhile, $Q$ MFGs are embedded inside those loops, implementing the current-to-flux conversion, as shown in Fig. 5(b).

The topology information of a MFG network can be described with only two matrices. The first one is the loop-to-loop inductance matrix $\mathbf{L_m}$, which describes the magnetic couplings of those $P$ loops, as illustrated in Fig. 5(c). The elements of this inductance matrix are self-inductance or mutual-inductance that can be extracted from the routings of wires. The second one is the loop-to-MFG incidence matrix $\boldsymbol{\sigma}$, which depicts the locations of MFGs inside those $P$ loops, as illustrated in Fig. 5(d).

For the Loop-$i$ ($1 \leq i \leq P$) with a loop current $i_{mi}$ circulating inside, its total coupled flux $\Phi_{mi}$ is contributed by loop currents, in addition to the external flux $\Phi_{ei}$. If we rewrite the Kirchhoff's voltage law (KVL) and the FQL with nominal fluxes, we find that $\Phi_{mi}$ in Loop-$i$ equals to the total flux contributions of MFGs inside the loop, no matter whether it is a normal or superconducting loop, namely

$$\Phi_{ei} - \sum_{j=1}^{P} l_{ij} i_{mj} = \Phi_{mi} = \sum_{j=1}^{Q} \sigma_{ij} \Phi_{ELj} + Const_i; (i=1,\cdots,P) \quad (2)$$

where $Const_i = n\Phi_0$ ($n$ is an integer) is the constant for a superconducting loop, and $Const_i = r\Phi_0$ ($r$ is a real number) is the constant for a normal loop.

For the MFG-$j$ ($1 \leq j \leq Q$), its branch current $i_{ELj}$ comes from the loops this MFG is inserted in. According to the equivalent circuit shown in Fig. 4(a), the current-to-flux function associated with the loop currents is derived as,

$$C_j \frac{d^2\Phi_{ELj}}{dt^2} + G_j \frac{d\Phi_{ELj}}{dt} + I_{0j} \sin\left(2\pi \frac{\Phi_{ELj}}{\Phi_0}\right) + i_{nj} - i_{bj}$$
$$= i_{ELj} = \sum_{i=1}^{P} \sigma_{ij} i_{mi}; (j=1,\cdots,Q) \quad (3)$$

where $I_{0j} = 0$ for a normal RC component, and $I_{0j} > 0$ for a Josephson junction.

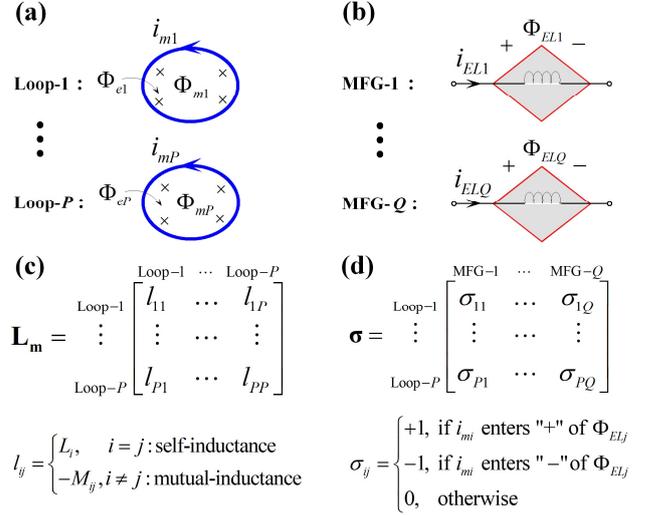

Fig. 5. Two objects and their topology information of a given SQUID circuit: (a) A set of $P$ loops, where a Loop-$i$ ($1 \leq i \leq P$) contains a total flux $\Phi_{mi}$ and a loop current $i_{mi}$ corresponding to the external flux $\Phi_{ei}$. (b) A group of $Q$ junctions, where an Element-$j$ ($1 \leq j \leq Q$) under a branch current $i_{ELj}$ makes a flux contribution $\Phi_{ELj}$ to loops. (c) The loop-to-loop inductance matrix $\mathbf{L_m}$, where the $L_i$ is the self-inductance of Loop-$i$, and $M_{ij}$ is the mutual-inductance between Loop-$i$ and Loop-$j$ ($1 \leq i,j \leq P$), the minus sign before $M_{ij}$ indicates that the flux contributed by the mutual-inductance is opposite to the one by the self-inductance. (d) The loop-to-MFG incidence matrix $\boldsymbol{\sigma}$, where $\sigma_{ij}$ describes the connection and polarity between $i_{mi}$ ($1 \leq i \leq P$) and $i_{ELj}$ ($1 \leq j \leq Q$).

### C. General Equation of MFG Networks

Combine the equations in (2) and (3), we can derive a network equation in matrix as,

$$\mathbf{C} \cdot \frac{d^2\boldsymbol{\Phi_{EL}}}{dt^2} + \mathbf{G} \cdot \frac{d\boldsymbol{\Phi_{EL}}}{dt} + \mathbf{I_0} \cdot \sin\left(\frac{2\pi\boldsymbol{\Phi_{EL}}}{\Phi_0}\right) + \boldsymbol{\sigma}^T \mathbf{L_m^{-1}} \boldsymbol{\sigma} \cdot \boldsymbol{\Phi_{EL}} \quad (4)$$
$$= \mathbf{i_b} - \mathbf{i_n} + \boldsymbol{\sigma}^T \mathbf{L_m^{-1}} \cdot (\boldsymbol{\Phi_e} - \mathbf{Const})$$

where, $\boldsymbol{\sigma}^T$ is the transpose of $\boldsymbol{\sigma}$, and $\mathbf{L_m^{-1}}$ is the inverse matrix of $\mathbf{L_m}$; $\mathbf{C}$, $\mathbf{G}$, $\mathbf{I_0}$ are parameter diagonal matrices of MFGs; $\boldsymbol{\Phi_{EL}}$ is the vector the flux output of MFGs; $\mathbf{i_b}$, $\boldsymbol{\Phi_e}$ and $\mathbf{Const}$ are the input vectors for MFGs and loops; they are defined as

$$\begin{cases} \mathbf{C} = diag\{C_1, \cdots, C_Q\} \\ \mathbf{G} = diag\{G_1, \cdots, G_Q\} \\ \mathbf{I_0} = diag\{I_{01}, \cdots, I_{0Q}\} \\ \boldsymbol{\Phi_{EL}} = [\Phi_{EL1} \cdots \Phi_{ELQ}] \\ \mathbf{i_b} = [i_{b1} \cdots i_{bQ}]; \mathbf{i_n} = [i_{n1} \cdots i_{nQ}] \\ \boldsymbol{\Phi_e} = [\Phi_{e1} \cdots \Phi_{eP}] \\ \mathbf{Const} = [Const_1 \cdots Const_P] \end{cases} \quad (5)$$

This network equation shows that the differences between Josephson junction circuits and normal RLC circuits are exhibited by the values of **Const** and **I₀**. Since **Const = 0** is suitable for both superconducting and non-superconducting loops, the sine function with **I₀ ≠ 0** in (4) is the special item for Josephson junction circuits. In other words, Josephson junction circuits are actually the RLC networks driven by the additional Josephson currents.

Therefore, to analysis and simulate a Josephson junction circuit, we only need to transform the given circuit into a MFG network, then exact the circuit components and parameters to set up the general network equation, according to Fig. 5. Finally, we can use one general solver of (4) to simulate the dynamics and characteristics of the circuit with numerical solutions.

### III. APPLICATION EXAMPLES

#### A. Rf-SQUID Analysis

From the MFG network of the rf-SQUID circuit shown in Fig. 10, we extract the parameters for the network equation as listed in Table I. For simplicity, all the all the circuit parameters are normalized with a reference resistor $R_0$ and a reference critical current $I_0$; for example, any inductance $L_x$ is normalized as $\beta_{Lx}$, $\beta_{Lx} = 2\pi I_0 L_x/\Phi_0$; any capacitance $C_x$ is normalized as $\beta_{Cx}$, $\beta_{Cx} = 2\pi I_0 R_0^2 C_x/\Phi_0$.

By solving the network equation configured with practical circuit parameters, typical current-voltage characteristics and flux-to-voltage curves are calculated, as shown in Fig. 6. They are periodically flux-modulated as we can observe from the experimental results [17].

TABLE I. PARAMETERS FOR A RF-SQUID CIRCUIT

| Parameter | Symbol | Value(Unit) |
|---|---|---|
| Number of Loops | $P$ | 2 |
| Number of MFGs | $Q$ | 2 |
| Reference current | $I_0$ | 10 (µA) |
| Reference resistance | $R_0$ | 1 (Ω) |
| Bias currents | $\mathbf{i_b}$ | $i_{b1} = 0$; $i_{b2} = I_{rf}\sin(2\pi f_{rf}t)$ $f_{rf} = 1/2\pi\sqrt{(L_1C_1)}$ |
| Flux inputs | $\mathbf{\Phi_e}$ | $\Phi_{e1} = \Phi_{in}$, $\Phi_{e2} = 0$ |
| Capacitance matrix | **C** | $\beta_{C1} = 2.0$; $\beta_{C2} = 0$ |
| Conductance matrix | **G** | $G_1 = R_0$; $G_2 = 1/(50R_0)$ |
| Critical current matrix | **I₀** | $I_{01} = I_0$, $I_{02} = 0$. |
| Inductance matrix | **L** | $\beta_{L1} = 2.0$; $\beta_{L2} = 100$; $\beta_{M12} = 8.0$ |
| Incidence matrix | **σ** | $\sigma_{11} = 1$; $\sigma_{22} = -1$; $\sigma_{12} = \sigma_{21} = 0$ |

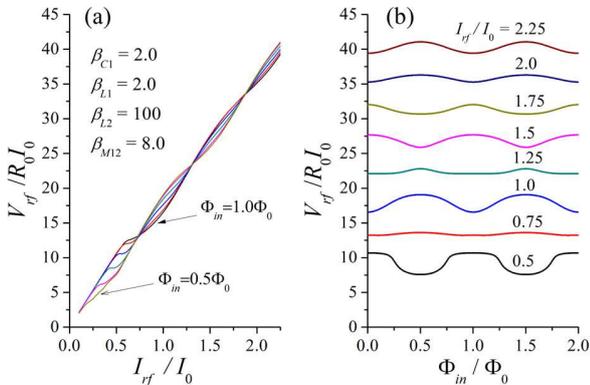

Fig. 6. Simulation results of the rf-SQUID circuit: (a) the flux-modulated current-voltage characteristics, and (b) the flux-to-voltage characteristics, where $V_{rf}$ is the root-mean-square (RMS) voltage of MFG-2, $V_{rf} = \langle (u_{EL2})^2 \rangle$.

#### B. Dc-SQUID Analysis

There are only one loop and two MFGs in the MFG diagram shown in Fig. 11. The parameters for the network equation of the dc-SQUID are extracted as in Table II. The current-voltage characteristics and the flux-to-voltage curves of the dc-SQUID are simulated, as shown in Fig. 7. They are same with the typical simulation results obtained with conventional analysis methods [18].

TABLE II. PARAMETERS FOR A DC-SQUID CIRCUIT

| Parameter | Symbol | Value(Unit) |
|---|---|---|
| Number of Loops | $P$ | 1 |
| Number of MFGs | $Q$ | 2 |
| Reference current | $I_0$ | 10 (µA) |
| Reference resistance | $R_0$ | 1 (Ω) |
| Bias currents | $\mathbf{i_b}$ | $i_{b1} = i_{b2} = I_b/2$ |
| Flux inputs | $\mathbf{\Phi_e}$ | $\Phi_{e1} = \Phi_{in}$ |
| Capacitance matrix | **C** | $\beta_{C1} = \beta_{C2} = 0.5$ |
| Conductance matrix | **G** | $G_1 = G_2 = 1/R_0$ |
| Critical current matrix | **I₀** | $I_{01} = I_{02} = I_0$ |
| Inductance matrix | **L** | $\beta_{L1} = 2.0$ |
| Incidence matrix | **σ** | $\sigma_{11} = 1$; $\sigma_{12} = -1$ |

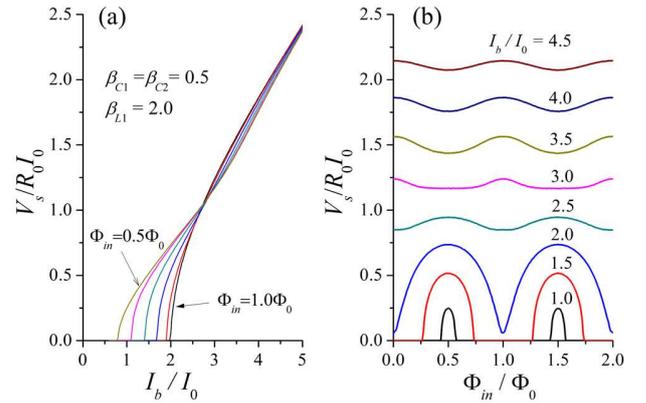

Fig. 7. Simulation results of the dc-SQUID circuit: (a) the current-voltage characteristics, and (b) the flux-to-voltage characteristics, where $V_s$ is the average voltage of MFG-1 or MFG-2, namely $V_s = \langle u_{EL1} \rangle$.

#### C. Bi-SQUID Analysis

From the MFG diagram shown in Fig. 12, the parameters of the network equation of bi-SQUID are extract in Table III.

TABLE III. PARAMETERS FOR A BI-SQUID CIRCUIT

| Parameter | Symbol | Value(Unit) |
|---|---|---|
| Number of Loops | $P$ | 2 |
| Number of MFGs | $Q$ | 3 |
| Reference current | $I_0$ | 10 (µA) |
| Reference resistance | $R_0$ | 1 (Ω) |
| Bias currents | $\mathbf{i_b}$ | $i_{b1} = i_{b2} = I_b/2$; $i_{b3} = 0$ |
| Flux inputs | $\mathbf{\Phi_e}$ | $\Phi_{e1} = \Phi_{in}$; $\Phi_{e2} = 0$ |
| Capacitance matrix | **C** | $\beta_{C1} = \beta_{C2} = \beta_{C3} = 0.5$ |
| Conductance matrix | **G** | $G_1 = G_2 = G_3 = 1/R_0$ |
| Critical current matrix | **I₀** | $I_{01} = I_{02} = I_0$; $I_{03} = 1.1 I_0$ |
| Inductance matrix | **L** | $\beta_{L1} = 2.0$; $\beta_{L2} = 0.1$; $\beta_{M12} = 0$ |
| Incidence matrix | **σ** | $\sigma_{11} = \sigma_{12} = 0$; $\sigma_{13} = \sigma_{21} = 1$; $\sigma_{22} = \sigma_{23} = -1$ |

The flux-to-voltage characteristics of the bi-SQUID are simulated as shown in Fig. 8, which agrees well with the typical results measured with the practical bi-SQUIDs [19].

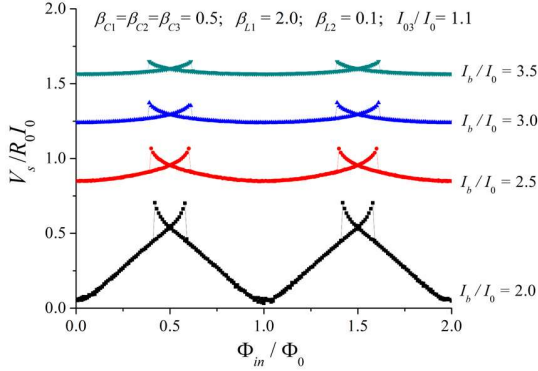

Fig. 8. The simulated flux-to-voltage characteristics of the bi-SQUID, where the voltage $V_s$ is the average voltage of MFG-1 or MFG-2, $V_s = \langle u_{EL1} \rangle$.

### D. Dro-SQUID Analysis

The MFG diagram of the dro-SQUID consists of two loops and four MFGs, as shown in Fig. 13. The parameters for its network equation are extracted as summarized in Table IV. The simulation results of are depicted in Fig. 9, where the typical flux-voltage curves of the dro-SQUID are square-wave shaped, as we can find in the typical experimental results [20].

TABLE IV. PARAMETERS FOR A dro-SQUID CIRCUIT

| Parameter | Symbol | Value(Unit) |
|---|---|---|
| Number of Loops | $P$ | 2 |
| Number of MFGs | $Q$ | 4 |
| Reference current | $I_0$ | 10 (μA) |
| Reference resistance | $R_0$ | 1 (Ω) |
| Bias currents | $\mathbf{i_b}$ | $i_{b1} = i_{b2} = i_{b3} = 0$; $i_{b4} = I_b$ |
| Flux inputs | $\mathbf{\Phi_e}$ | $\Phi_{e1} = \Phi_{in}$; $\Phi_{e2} = 0$ |
| Capacitance matrix | $\mathbf{C}$ | $\beta_{C1} = \beta_{C2} = 0.2$; $\beta_{C3} = 0.3$; $\beta_{C4} = 0$ |
| Conductance matrix | $\mathbf{G}$ | $G_1 = G_2 = G_3 = 1/(5R_0)$; $G_4 = 1/R_0$ |
| Critical current matrix | $\mathbf{I_0}$ | $I_{01} = I_{02} = I_0$; $I_{03} = 1.5 I_0$; $I_{04} = 0$ |
| Inductance matrix | $\mathbf{L}$ | $\beta_{L1} = 1.7$; $\beta_{L2} = 2.0$; $\beta_{M12} = 0.8$ |
| Incidence matrix | $\mathbf{\sigma}$ | $\sigma_{11} = \sigma_{22} = \sigma_{23} = 1$; $\sigma_{12} = \sigma_{24} = -1$; $\sigma_{13} = \sigma_{14} = \sigma_{21} = 0$ |

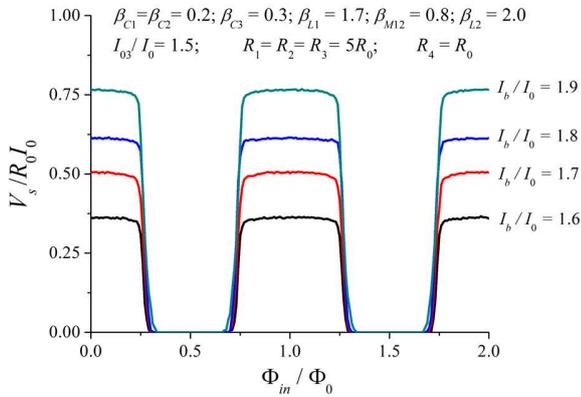

Fig. 9. The flux-to-voltage characteristics of the dro-SQUID simulated through solving the general network equation, where the $V_s$ is the average voltage of MFG-3, $V_s = \langle u_{EL3} \rangle$.

## IV. DISCUSSION AND CONCLUSION

We introduced a MFG model to realize the integration of Josephson junction circuits and normal RLC circuits, and derived a unified network equation to simplify their dynamic analysis and simulation. The MFG concept and the network equation provide three insights for the design and analysis Josephson junction circuits, which solve the problems encountered in the nodal analysis method, although two methods are equivalent in circuit theories.

(1) Josephson junctions are the Josephson-current-shunted RC components, and Josephson junction circuits are the RLC networks modified by Josephson currents. The sine item in (4) is the factor that makes Josephson junction circuits different from the normal RLC networks.

(2) Normal and superconducting RLC circuits are both the MFG networks. They are the dynamic system of MFGs and loops, where, MFGs are generating magnetic fluxes into loops, while the magnetically-coupled loops convert the coupled fluxes into loop currents to drive the MFGs.

(3) The $\Phi_{EL}$ of a MFG records the total flux contribution of the MFG, and $u_{EL}$ is correspondingly the flow rate of $\Phi_{EL}$. The circuit variables of MFG networks have physical meanings for both superconducting and normal elements, and can be easily understood by the electronic engineers.

The advantage of this network equation is that the inductance matrix $\mathbf{L_m}$ includes all the magnetic couplings between loops; the simulation based on this equation is precise if we can build a tool to extract the inductance matrix $\mathbf{L_m}$ directly from the circuit layout. Meanwhile, the network equation is a group of second-order ordinary differential equations that can be efficiently processed using one general solver. Therefore, this unified network equation is promising to develop a precise and efficient simulation tool for Josephson junction circuits and their hybrids with normal RLC elements.

## APPENDIX

### A. Rf-SQUID

In the circuit diagram shown in Fig.2(a), we can define two MFGs, as shown in Fig. 11(a). Josephson junction $J_1$ without the bias current source is defined as MFG-1; the combination of the rf current source $i_{rf}(t)$ and the 50Ω impedance is modeled as MFG-2.

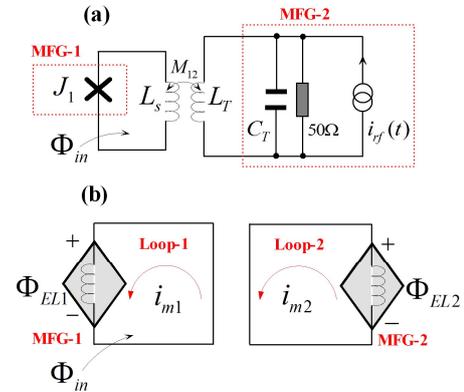

Fig. 10. (a) MFGs defined in the rf-SQUID circuit, and (b) the equivalent MFG network, where the inductances of loops are not shown for simplicity.

Accordingly, the rf-SQUID circuit is transformed into a MFG network, where MFG-1 and MFG-2 generate fluxes into Loop-1 and Loop-2, respectively; they are interfering each other through the mutual-inductance $M_{12}$, as shown in Fig. 10(b).

*B. Dc-SQUID*

The dc-SQUID circuit shown in Fig. 2(b) is redrawn as shown in Fig. 11(a), where two Josephson junctions are defined as MFG-1 and MFG-2, and the current source $I_b$ is divided equally for two MFGs. The MFG network of the dc-SQUID has two MFGs interfering each other in one loop, as shown in Fig. 11(b).

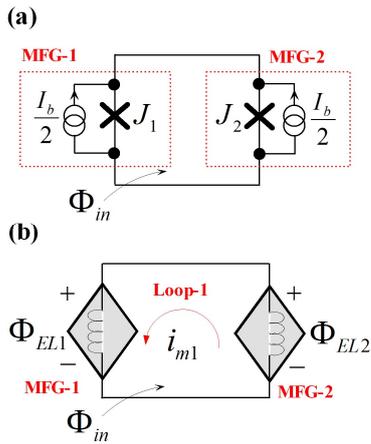

Fig. 11. (a) MFGs defined in the dc-SQUID circuit, and (b) the equivalent MFG network of the dc-SQUID circuit.

*C. Bi-SQUID*

The bi-SQUID circuit shown in Fig. 2(c) contains three MFGs, as shown in Fig. 12(a), where the current source $I_b$ is separately assigned to MFG-1 and MFG-2. Its MFG diagram is shown in Fig. 12(b), where three MFGs form two loops.

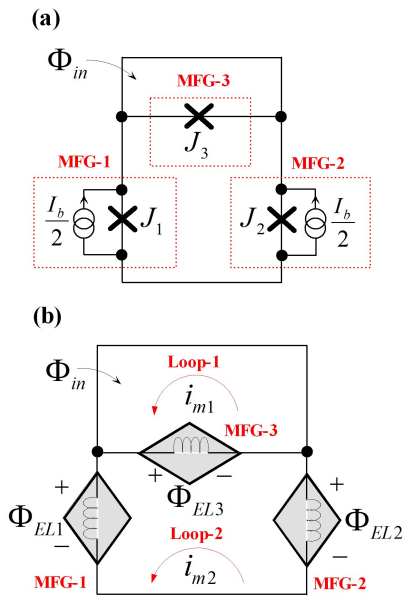

Fig. 12. (a) MFGs defined in the bi-SQUID circuit. (b) The equivalent MFG network of the bi-SQUID.

*D. Dro-SQUID*

The dro-SQUID circuit shown in Fig. 2(d) is redrawn as shown in Fig. 13(a), where three Josephson junctions are corresponding to three MFGs, while the combination of the bias current source $I_b$ and the shunt resistor $R_a$ is modeled as the MFG-4. The MFG diagram of the dro-SQUID is composed of four MFGs and two loops, as shown in Fig. 13(b).

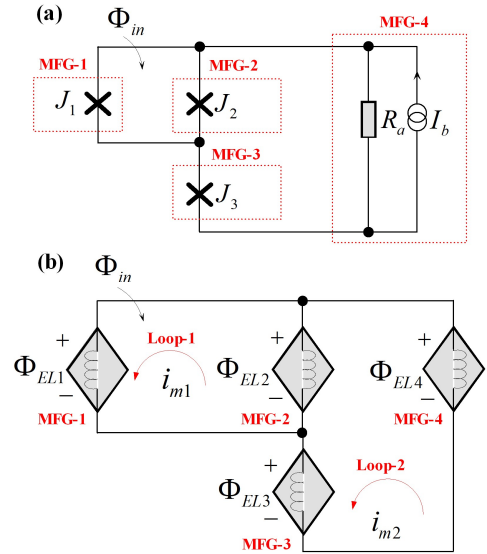

Fig. 13. (a) MFGs defined in the dro-SQUID circuit. (b) The equivalent MFG network of the dro-SQUID circuit.